\documentstyle[aps,epsf,prl,twocolumn]{revtex}

\draft

\begin{document}

\title{Domain Coarsening in Electroconvection}

\author{Lynne Purvis}
\author{Michael Dennin}
\address{Department of Physics and Astronomy}
\address{University of California at Irvine}
\address{Irvine, CA 92697-4575.}

\date{\today}

\maketitle

\begin{abstract}

We report on experimental measurements of the growth of regular
domains evolving from an irregular pattern in electroconvection.
The late-time growth of the domains is consistent with the
size of the domains scaling as $t^n$. We use two isotropic
measurements of the domain size: the structure factor
and the domain wall length. Measurements using
the structure factor are consistent with $t^{1/5}$ growth.
Measurements using the
domain wall length are consistent with $t^{1/4}$ growth.
One source of this discrepancy is the fact that the distribution
of local wavenumbers is approximately independent of the domain size.
In addition, we measure the anisotropy of the growing domains.   
\end{abstract}


There are many situations where a system experiences a
rapid change of an external parameter, or quench, such
that the state of the system after the quench is not an
equilibrium or steady-state phase.
Domains of the new equilibrium phase form, and
the subsequent growth of these domains, or {\it coarsening},
is often characterized by a single scale size for the
domains that follows a power-law growth. There has been a great deal
of both theoretical and experimental study of this process for
systems where both the initial state prior to the quench
and final state after the quench are thermodynamic equilibrium
states at finite temperature \cite{REV}. We are
interested in the analogous process for systems
that are driven out of equilibrium. For such systems, neither the
initial nor the final state is in thermodynamic equilibrium; however,
they are steady-states of the system.
The first situation will be referred to as a
{\it thermodynamic system} and the second situation as
a {\it driven system}.
For driven systems, studies of model
equations suggest power-law scaling of the late-time domain
growth; however, the value of the growth exponent
depends on the measurement scheme \cite{EVG92,CM95,CB98}.
Experimentally, the coarsening of random patterns after
a ramp in the control pattern has been studied, but growth
exponents for domain size were not measured \cite{MAC91}. Thermodynamic
systems with stripes have been studied experimentally
using block copolymers \cite{HAC00}.

In spatially extended systems that are driven out of equilibrium,
there is generally a transition from a uniform state to a periodic, or
``stripe'', state at a critical value $R_c$
of the external control parameter $R$ \cite{PATTREV}.
A quench
corresponds to a rapid change of $R$. For values reasonably close
to $R_c$, these systems are often well described by model
equations that are similar to, and in some cases identical to,
the equations used to study coarsening in thermodynamic
systems \cite{PATTREV}.
However, in a driven system,
there is generally no equivalent of a free-energy that governs the 
growth of the domains, and the periodic structure
complicates the dynamics. 
Therefore, the question of how the ordering proceeds in driven
systems, and the differences and similarities
with thermodynamic systems, is one of great interest.

Simulations of potential \cite{CM95,CB98} and
nonpotential \cite{CM95} forms of
the Swift-Hohenberg equation
have been used to study quenches in
driven systems \cite{EVG92,CM95,CB98}.
In both cases, characterization of the growth by the
structure function
$S(k)$ suggests a length scale that grows as
$t^{1/5}$ \cite{CM95,CB98}.
In contrast, the growth exponent determined from the
orientational correlation function
is consistent with $1/4$ \cite{CM95,CB98} for potential dynamics
and with $1/2$ \cite{CM95} for nonpotential
dynamics. There is
no explanation of the discrepancy between the $S(k)$ and
orientational correlation function measurements;
however, the length scales determined by these measures
do not have the same immediate interpretation \cite{CB98}.
The orientational correlation function results agree
with experiments in block copolymers \cite{HAC00}.

We have made experimental measurements
of coarsening in an anisotropic, driven system:
electroconvection in a
nematic liquid crystal \cite{ECART}.
A nematic liquid crystal is a fluid in which the
molecules align on average along
a particular axis,
referred to as the director \cite{LC}, which we take as the x-axis.
Because the system is anisotropic, there are
only two possible orientations of the domains. This is fundamentally
different than previous simulations and experiments. One
expects a different scale size parallel and perpendicular to the preferred
direction in the system. Our measurement of these scale sizes show that
not only are the magnitudes of the length scales different, but they also
coarsen with different growth exponents. We also measure
isotropic properties of the domain growth and find that they are
in surprising agreement with the predictions of the Swift-Hohenberg
simulations \cite{CM95,CB98}. Finally, our measurements show that
the domain size and wavenumber variation have different growth
exponents, which explains the discrepancy between the
$S(k)$ and orientational correlation function measurements.

For electroconvection, the liquid crystal is placed between
two glass plates with the director aligned parallel to the plates and
along a single axis. The liquid crystal is doped with ionic impurities.
An ac voltage is applied perpendicular to the
plates. Above a critical value of the applied voltage $V_c$,
there is a transition to
a state that consists of convection rolls with a corresponding periodic
variation of the director and charge density.
We studied {\it oblique rolls}, i.e. patterns that have a nonzero
angle $\theta$ between the wavevector and the alignment of the undistorted
director. Oblique rolls with the same wavenumber $k$ but at $\theta$ (zig rolls)
and $-\theta$ (zag rolls) are degenerate.
The initial transition in this system is to a pattern
that consists of the superposition of four modes: right- and left-traveling
zig and zag rolls. The interaction of these four modes leads
to irregular spatial and temporal variations of the amplitudes of each
of the modes \cite{DAC96}, i.e. spatiotemporal chaos
\cite{PATTREV,EXP}. For traveling rolls,
a sufficiently large modulation of
the amplitude of the applied voltage at twice the intrinsic
frequency of the pattern stabilizes standing rolls \cite{RCK88,W88,RRFJS88}.
For our system, either standing zig or
standing zag rolls are
stabilized, and the stabilized pattern exhibits regular
temporal dynamics \cite{D00a}. In our experiments, a 
quench corresponded
to a rapid change of the modulation amplitude at a fixed
value of $V_{rms}$.
As standing waves can be stabilized both below and
above $V_c$, two types of quenches are possible.
Below $V_c$, the dynamics of the standing
waves are potential \cite{R90}, and above $V_c$,
they are nonpotential \cite{R00}. Therefore, based on the
results of Ref.~\cite{CM95}, we expected two different
growth exponents. Also, below $V_c$,
domains of standing zig and zag rolls must first
form before the system coarsens, as the initial state is
spatially uniform. This process is illustrated in Fig. 1(a) - (d).
Above $V_c$, domains
of zig and zag rolls are already present after the quench,
and the coarsening proceeds
from this initial distribution.
This process is illustrated in Fig. 1(e) - (h).

The details of the experimental apparatus are described in Ref.~\cite{D00a}.
Commercial cells \cite{EHCO} with a thickness
of 23~$\mu$m and 1~cm$^2$ electrodes
were used. The sample temperature was maintened at
$50.0 \pm 0.002\ {\rm ^{\circ}C}$. The patterns were observed from
above using
a modified shadowgraph setup \cite{ASR99} that
effectively eliminated the well-known nonlinear effects
in the shadowgraph image \cite{RHWR89}.
An ac voltage, $V(t) = [V_o + V_m\sin(\omega_m t)]\sin(\omega_d t)$,
was applied across the sample, with $\omega_d/2\pi = 25\ {\rm Hz}$.
There are two relevant dimensionless control
parameters: $\epsilon = (V_o/V_c)^2 - 1 $ and $b = V_m/V_c$.
Here $V_c$ is the critical voltage for the onset of a pattern
when $V_m = 0$. There was a small, linear
drift in $V_c$; therefore, before and after
each quench, $V_c$ and $\omega_h$ were determined. Typical
values were $V_c = 15\ {\rm V_{rms}}$ and $\omega_h/2\pi = 0.5\ {\rm Hz}$.

The system was equilibrated at $b = 0$ and
either $\epsilon = -0.03$ or $\epsilon = 0.03$ for five minutes.
For $\epsilon = -0.03$ ($\epsilon = 0.03$),
the transition to standing waves occurs
at $b = 0.03$ ($b = 0.013$) \cite{D00a}.
For $\epsilon = -0.03$, we used a jump from $b = 0$ to $b = 0.05$, and for
$\epsilon = 0.03$, we used a jump from $b = 0$ to $b = 0.04$. In both
cases, $\omega_m = 2\omega_h$, where $\omega_h$ is the frequency of
the pattern at onset, the Hopf frequency. A third 
quench was done at $\epsilon = 0.03$ with $\omega_m = 2.1 \omega_h$.
Immediately after a jump in $b$, a series of 128 images was obtained.
An image was taken every $5^{th}$ cycle
of the modulation, or roughly once every 5 seconds, triggered by the
applied voltage to occur at $T/4$, where $T$ is the period of
the modulation.
In order to resolve the individual rolls, a 1.35~mm
by 1.35~mm section of the sample was imaged.
At the end of a typical time
series, the domain size was on the order of our observation window; however,
given the size of the entire sample,
we were not limited by finite size effects. Unless
otherwise noted, time is scaled by the
director relaxation time, which for our system is 0.2~s.
For $\epsilon = 0.03$, we performed 20 quenches, and for
$\epsilon = -0.03$, we performed 18 quenches.
The domain wall length, $S(k)$,
the local wavenumber distribution,
$\Delta q_x$, and $\Delta q_y$ were computed for each
image in the time series for a given quench. ($\Delta q_x$
and $\Delta q_y$ are defined below.) 
The values at each time step
were averaged over all the quenches for a given
$\epsilon$. These averaged values were used to compute the growth
exponents, which did
not change significantly after averaging 10 quenches.

If a single scale length is sufficient to describe the domain growth,
both the total domain wall length and the width of the power
spectrum would exhibit the same growth exponent.
Because, the
structure factor contains information about the range of
local wavenumbers within a domain, we measured
the local wavenumber using the method described in Ref.~\cite{EMB98}.
Briefly, this method involves
calculating the x-component of $k$ from
$|k_x|^2 = -[\partial_x^2 u({\bf x})]/u({\bf x})$,
with a similar calculation for the y-component. Here $u({\bf x})$ is
the image of the pattern. We used
the central 0.82~mm by 0.82~mm region of the image in the
determination of the spread $\Delta k$ of the magnitude
$k$ of the local
wavevector ${\bf k}$. We defined $\Delta k$ as
the square root of the second moment of
the distribution of $k$ about $k_{ave}$.

To determine the normalized domain-wall length $L$, the local
orientation was determined from the sign of
$[\partial_{xy} u({\bf x})]/u({\bf x})$. Regions
with a negative (positive)
value of this ratio corresponded to zig (zag)
rolls and were assigned a grayscale of 0 (255).
The resulting image was smoothed, producing
an image in which domain walls
had a value close to 128. Applying a threshold produced images
in which
zig regions had a value of 0, domain walls a value of 128, and
zag rolls a value of 255. The domain wall ``length'' was taken as
the total number of pixels of value 128 normalized by the
total number of pixels in the image. Images from a typical
sequence, after processing, are shown in Fig. 1.
The processed images were also used to study the
anisotropy of the growth. Since the wavelength has been factored
out of the processed images, the width of the central peak in the
power spectrum $S({\bf q})$ of one of
these images corresponds to the length scale of the domains.
Defining $\Delta q_x$ ($\Delta q_y$) as the second moment of
$q_x$ ($q_y$) about ${\bf q} = 0$ of $S({\bf q})$, with $q_x$ and
$q_y$ the x- and y-components of ${\bf q}$, provides
a measure of the inverse of the correlation
length in the x and y directions.

\begin{figure}[htb]
\epsfxsize = 3.0in
\epsfbox{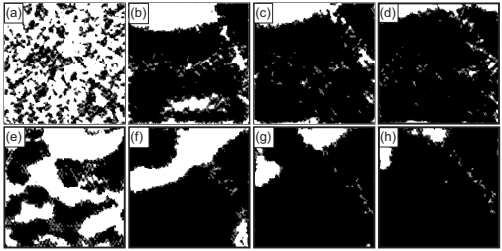}
\caption{Processed images illustrating the domain
growth for a quench at $\epsilon = -0.03$ (images (a) - (d))
and for a quench at $\epsilon = 0.03$ (images (e) - (h)).
The images cover an area of 1.35~mm by 1.35~mm, and the processing
method is described in detail in the text. The black regions
are areas of zig rolls, and the white areas are regions of
zag rolls. Image (a)/(f) was taken 5~s after the quench, and
the subsequent images are all 160 s apart.}
\end{figure}

The measurements based on $S(k)$ used the
same method as described in Ref.~\cite{CM95}. The structure
factor (the square of the Fourier transform) was averaged over
all angles. The relevant peak in the
resulting $S(k)$ was fit to a Lorentzian squared,
and the width $\delta k$
was defined as the half width at half height.
This method is an isotropic
measure of the domain growth and is used to ensure
that our results are
directly comparable to Ref~\cite{CM95}.

\begin{figure}[htb]
\epsfxsize = 3.5in
\epsfbox{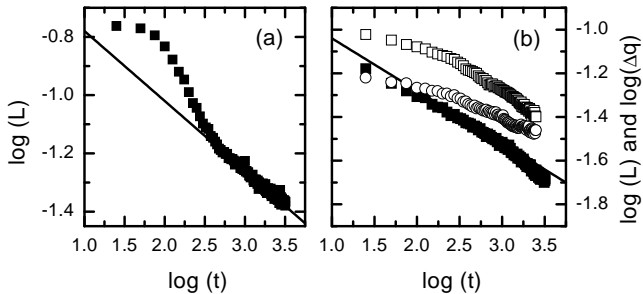}
\caption{This shows a plot of $\log(L)$ versus $\log(t)$. Here
$L$ is the total length of the domain wall in the region of
study.
Plot (a) is for the quench at $\epsilon = -0.03$,
and plot (b) is for the quench at $\epsilon = 0.03$. The solid lines
are linear fits to the data. The line in (a) has a slope
of $-0.240 \pm 0.004$, and the line in (b) has a slope of
$-0.243 \pm 0.003$. In (b), the open symbols are the results for
$\Delta q_x$ (circles) and $\Delta q_y$ (squares).}
\end{figure}

Figures~2a and 2b show the results for the domain
wall length $L$ for the $\epsilon = -0.03$ and $\epsilon = 0.03$
quenches, respectively. Here $\log(L)$ is plotted versus
$\log(t)$. For the quench at
$\epsilon = 0.03$, the behavior is consistent with power law
scaling essentially immediately after the quench. For
$\epsilon = -0.03$, the system is not consistent with
power law scaling until $\log(t) \approx 2.5$.
This difference is reasonable given that the domains
must first form for the quench at $\epsilon = -0.03$. Also,
the scaling occurs in both systems at roughly the
same scale size for the domains, $\log(L) \approx -1.2$.
The decay of the domain-wall
length is consistent with scaling
as $t^{-1/4}$.
The solid line in Fig.~2a is a fit
over the range $2.5 < \log(t) < 3.5$ and has a slope of
$-0.24$. The solid line in Fig.~2b is a fit
over the range $2.0 < \log(t) < 3.2$ and also has a slope of $-0.24$.

Figure 2b also shows the results for $\Delta q_x$ (open circles)
and $\Delta q_y$ (open squares). As is suggested by the images in
Fig.~1, the domains tend to be larger along the director (x-direction).
This is confirmed by the relative magnitudes of $\Delta q_x$ and
$\Delta q_y$. Also, the two lengths coarsen with different
exponents, where the growth exponent perpendicular to the director is
consistent with the exponent measured using the domain-wall length.
A similar result holds for the $\epsilon = -0.03$ quench, with the
same delay in the onset of scaling as seen in the wall length.

\begin{figure}[htb]
\epsfxsize = 3.5in
\epsfysize = 2.5in
\epsfbox{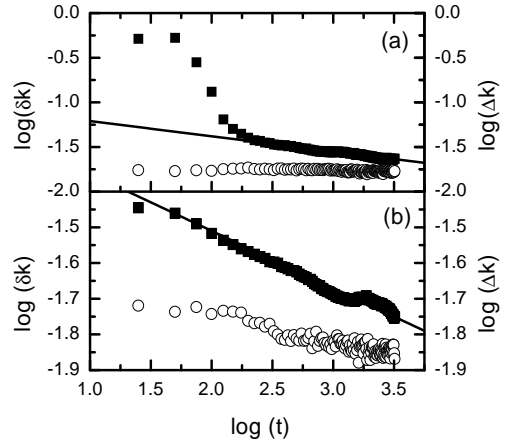}
\caption{This shows a plot of $\log(\delta k)$ versus $\log(t)$ 
using the left hand axis and solid symbols, and
a plot of $\log(\Delta k)$ versus $\log(t)$ using the right hand
axis and open symbols. Plot (a) is for the quench at $\epsilon = -0.03$,
and plot (b) is for the quench at $\epsilon = 0.03$. The solid lines
are linear fits to the solid data. The line in (a) has a slope
of $-0.170 \pm 0.003$, and the line in (b) has a slope of
$-0.165 \pm 0.002$.}
\end{figure}

Figures~3a and 3b show $\log(\delta k)$ and
$\log(\Delta k)$ versus $\log(t)$ for the
quenches at $\epsilon = -0.03$ and $\epsilon = 0.03$,
respectively.
The behavior is
consistent with power-law scaling almost immediately
for $\epsilon = 0.03$ and at later times for $\epsilon = -0.03$.
The scaling sets in for both
quenches at approximately the same domain size.
For both quenches, the scaling of
$\delta k (t)$ is consistent with
$t^{-1/5}$ decay, as found in simulations \cite{CM95,CB98}.
For $\epsilon = -0.03$, the solid line
is a fit over the region $2.5 < \log(t) < 3.5$ and
has a slope of -0.17. For $\epsilon = 0.03$, the solid line is
a fit over the region $2.0 < \log(t) < 3.2$ and has
a slope of -0.16. However, for both cases, the variation in the
local wavenumber $\Delta k$ is a significant fraction
of $\delta k$ in the possible scaling regime and is roughly constant
in time. Therefore,
$\delta k$ is not an accurate measure of the growth of the domain
size, but a complicated convolution
of the domain size and $\Delta k$. Accounting
for this effect, measurements of the coarsening parallel and
perpendicular to the director based on individual peaks of
$S({\bf k})$ are consistent with the anisotropy
determined from $S({\bf q})$ of the processed images.

Despite the different symmetries, the
isotropic measures of the domain growth in electroconvection agree well
with the simulations of the Swift-Hohenberg model \cite{EVG92,CM95,CB98}
and experiments in block copolymers \cite{HAC00}.
Understanding this agreement will require further
work. In particular, the work with block copolymers
attributes the exponent of $1/4$ to the dynamics of topological
defects $\cite{HAC00}$. Because of the anisotropy, the relevant
defect dynamics in electroconvection are different.
For example,
there are no regions where the roll orientation changes continuously,
only sharp domain walls.

\begin{figure}[t!]
\epsfxsize = 3.5in
\epsfysize = 2.0 in
\epsfbox{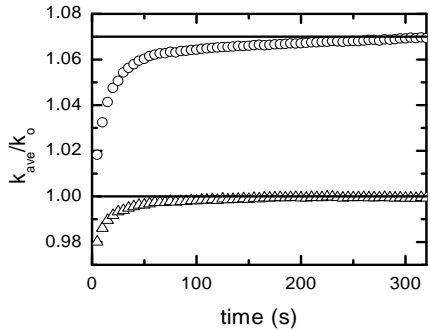}
\caption{This shows a plot of $k_{ave}/k_o$ versus time. 
The triangles are for the quench at $\epsilon = 0.03$ with
$\omega_m = 2.0\omega_h$, and the circles are for the
quench at $\epsilon = 0.03$ with $\omega_m = 2.1\omega_h$.
The lines are a guide to the eye.}
\end{figure}

One discrepancy is
our observation of a growth exponent of
$1/4$ for both $\epsilon < 0$ and $\epsilon > 0$.
For $\epsilon > 0$,
we expected an exponent of $1/2$ \cite{CM95}.
One possible explanation of
the discrepancy is the potential existence of long-time
transients. An analysis of the Swift-Hohenberg
equation for patterns with $k = k_o$ suggests a long-time
transient regime for which
the growth exponent is
$1/4$ \cite{EVG92,CB98}. In our system, the dispersion relation
fixes the average wavenumber
$k_{ave} = k_{o}$ for $\omega_m = 2\omega_h$. To vary $k_{ave}$,
we performed
a quench at $\omega_m = 2.1\omega_h$, for which $k_{ave} = 1.07k_o$.
The quench was at $\epsilon = 0.03$ to a value of $b = 0.04$.
Measurements of $L$ were consistent with
a growth exponent of $1/4$.
However, the average wavenumber exhibited a slow approach to
a steady state (see Fig. 4), suggesting that this
growth exponent also corresponds to a transient regime. 
Because there is no significant evolution of the wavenumber
spread, we plan to study larger regions of
the sample, without resolving
the individual rolls, for longer times by exploiting optical properties
of the patterns. This will allow for the
possibility of observing a crossover to a different growth exponent.

The authors acknowledge funding from NSF grant DMR-9975479.
L. Purvis was supported by NSF Research Experience for
Undergraduates grant PHY-9988066. M. Dennin would like to
thank Michael Cross and Herman Riecke for useful discussions.

\end{document}